\documentclass[useAMS,usenatbib]{mn2e}
\usepackage{times}
\usepackage{graphicx}
\title[The metamorphosis of ASASSN-15ed from SN Ibn to SN Ib]{Massive stars exploding in a He-rich circumstellar medium - VII. The metamorphosis of ASASSN-15ed from a narrow line Type Ibn to a normal Type Ib Supernova}
\author[Pastorello et al.]{A. Pastorello$^{1}$\thanks{E-mail: andrea.pastorello@oapd.inaf.it}, J. L. Prieto$^{2,3}$, N. Elias-Rosa$^{1}$, D. Bersier$^{4}$, G. Hosseinzadeh$^{5,6}$,
\newauthor A. Morales-Garoffolo$^{7}$, U. M. Noebauer$^{8}$, S. Taubenberger$^{8,9}$, L. Tomasella$^{1}$,
\newauthor   C. S. Kochanek$^{10,11}$, E. Falco$^{12}$, U. Basu$^{10,13}$, J. F. Beacom$^{10,11,14}$, S. Benetti$^{1}$, 
\newauthor   J. Brimacombe$^{15}$, E. Cappellaro$^{1}$, A. B. Danilet$^{14}$, Subo Dong$^{16}$, J. M. Fernandez$^{17}$,  
\newauthor   N. Goss$^{10,13}$, V. Granata$^{1,18}$, A. Harutyunyan$^{19}$, T. W.-S. Holoien$^{10}$, E. E. O. Ishida$^{8}$,
\newauthor   S. Kiyota$^{20}$, G. Krannich$^{21}$, B. Nicholls$^{22}$,  P. Ochner$^{1}$, G. Pojma\'nski$^{23}$,
\newauthor  B. J. Shappee$^{24,25}$, G. V. Simonian$^{10}$, K. Z. Stanek$^{10,11}$, S. Starrfield$^{26}$, D. Szczygie{\l}$^{23}$,
\newauthor   L. Tartaglia$^{1,18}$, G. Terreran$^{1,27}$, T. A. Thompson$^{10,11}$, M. Turatto$^{1}$, R. M. Wagner$^{10,28}$, 
\newauthor  W. S. Wiethoff$^{29,30}$, A. Wilber$^{26}$, P. R. Wo\'zniak$^{31}$.
\\
$^{1}$INAF –- Osservatorio Astronomico di Padova, Vicolo dell'Osservatorio 5, I-35122 Padova, Italy\\
$^{2}$N\'ucleo de Astronom\'ia de la Facultad de Ingenier\'ia, Universidad Diego Portales, Av. Ej\'ercito 441, Santiago, Chile\\
$^{3}$Millennium Institute of Astrophysics, Santiago, Chile\\
$^{4}$Astrophysics Research Institute, Liverpool John Moores University, 146 Brownlow Hill, Liverpool L3 5RF, UK\\
$^{5}$Las Cumbres Observatory Global Telescope Network, 6740 Cortona Dr., Suite 102, Goleta, CA 93117, USA\\
$^{6}$Department of Physics, University of California, Santa Barbara, Broida Hall, Mail Code 9530, Santa Barbara, CA 93106-9530, USA\\
$^{7}$Institut de Ci\`encies de l'Espai (CSIC-IEEC), Campus UAB, Carrer de Can Magrans S/N, E-08193 Cerdanyola del Vall\`es, Spain\\
$^{8}$Max-Planck-Institut f\"ur Astrophysik, Karl-Schwarzschild-Str. 1, D-85748 Garching, Germany\\
$^{9}$European Organisation for Astronomical Research in the Southern Hemisphere (ESO), Karl-Schwarzschild-Str. 2, D-85748 \\Garching bei M\"unchen, Germany\\
$^{10}$Department of Astronomy, The Ohio State University, 140 West 18th Avenue, Columbus, OH 43210, USA\\
$^{11}$Center for Cosmology and AstroParticle Physics (CCAPP), The Ohio State University, 191 W. Woodruff Ave., Columbus, OH 43210, USA \\
$^{12}$Harvard-Smithsonian Center for Astrophysics, Cambridge, MA 02138, USA\\
$^{13}$Grove City High School, 4665 Hoover Road, Grove City, OH 43123, USA\\
$^{14}$Department of Physics, The Ohio State University, 191 W. Woodruff Ave., Columbus, OH 43210, USA\\
$^{15}$Coral Towers Observatory, Unit 38 Coral Towers, 255 Esplanade, Cairns 4870, Australia\\
$^{16}$Kavli Institute for Astronomy and Astrophysics, Peking University, Yi He Yuan Road 5, Hai Dian District, Beijing 100871, China \\
$^{17}$Observatory Inmaculada del Molino, Hernando de Esturmio 46, E-41640 Osuna, Spain\\
$^{18}$Dipartimento di Fisica e Astronomia, Universit\`a degli Studi di Padova, Vicolo dell'Osservatorio 3, I-35122 Padova, Italy\\
$^{19}$Fundaci\'on Galileo Galilei-INAF, Telescopio Nazionale Galileo, Rambla Jos Ana Fern\'andez P\'erez 7, E-38712 Bre\~na Baja, TF, Spain\\
$^{20}$Variable Star Observers League in Japan (VSOLJ), 7-1 Kitahatsutomi, Kamagaya, Chiba 273-0126, Japan\\
$^{21}$Roof Observatory Kaufering, Lessingstr. 16, D-86916 Kaufering, Germany\\
$^{22}$Mount Vernon Observatory, 6 Mt. Vernon PI, Nelson 7010, New Zealand\\
$^{23}$Warsaw University Astronomical Observatory, Al. Ujazdowskie 4, PL-00-478 Warsaw, Poland\\
$^{24}$Hubble Fellow\\
$^{25}$Observatories of the Carnegie Institution for Science, 813 Santa Barbara Street, Pasadena, CA  91101, USA\\
$^{26}$School of Earth and Space Exploration, Arizona State University, Tempe, AZ 85287-1404, USA\\
$^{27}$Astrophysics Research Centre, School of Mathematics and Physics, Queen's University Belfast, Belfast BT7 1NN, UK\\
$^{28}$Large Binocular Telescope Observatory, University of Arizona, 933 N. Cherry Ave., Tucson, AZ 85721-0065, USA \\
$^{29}$SOLO Observatory, 85095 Ravine Road, Port Wing, WI 54865, USA\\
$^{30}$Department of Geological Sciences, 1049 University Drive, University of Minnesota, Duluth, MN 55812, USA\\
$^{31}$Los Alamos National Laboratory, Los Alamos, NM 87545, USA
}

\begin{document}

\date{Accepted XXXX Months XX. Received XXXX Month XX; in original form XXXX Month XX}

\pagerange{\pageref{firstpage}--\pageref{lastpage}} \pubyear{2015}

\maketitle

\label{firstpage}

\clearpage

\begin{abstract}
We present the results of the spectroscopic and photometric  monitoring campaign of ASASSN-15ed.
The transient was discovered quite young by the All Sky Automated Survey for SuperNovae (ASAS-SN). Amateur astronomers allowed us
to sample the photometric SN evolution around maximum light, which we estimate to have occurred on
JD = 2457087.4 $\pm$ 0.6 in the $r$-band. Its apparent $r$-band magnitude at maximum was $r$ = 16.91 $\pm$ 0.10, 
providing an absolute magnitude $M_r$ $\approx -20.04 \pm 0.20$, which is slightly more luminous than 
the typical magnitudes estimated for Type Ibn SNe. 
The post-peak evolution was  well monitored, and the decline rate (being in most bands around 0.1 mag d$^{-1}$ during 
the first 25 d after maximum) 
is marginally slower than the average decline rates of SNe Ibn during the same time interval. 
The object was initially classified as a Type Ibn SN because early-time spectra were characterized
by a blue continuum with superimposed narrow P-Cygni lines of He I, suggesting the presence of
a slowly moving (1200-1500  km s$^{-1}$), He-rich circumstellar medium. Later on, broad P-Cygni He I lines
became prominent. The inferred velocities, as measured from the  minimum of the broad absorption
components, were between 6000 and 7000 km s$^{-1}$. As we attribute these broad features to  the SN ejecta,
this is the first time we have observed the transition of a Type Ibn SN to a Type Ib SN. 
\end{abstract}


\section{Introduction}

Type Ibn supernovae (SNe) are a class of stripped-envelope SNe characterized by evidence of interaction 
between the SN ejecta and a He-rich and H-depleted circumstellar medium (CSM). Although the first SN of this class 
was discovered at the end of twentieth century \citep[SN 1999cq,][]{mat00}, the first well-sampled 
example was SN 2006jc \citep[][and references therein]{pasto08a}. SN 2006jc
is a sort of milestone in the studies of SNe Ibn, since it occurred in a nearby galaxy, UGC 4904 (d $\approx$ 25 Mpc),
was observed over a wide range of wavelengths, and was the very first object for which a pre-SN eruption
was observed \citep{yam06,fol07,pasto07}.

So far, about 20 objects classified as Type Ibn SNe have been discovered, showing a wide variety in their 
observed properties. Some transitional SNe Ibn show evidence of H spectral lines produced in their CSM \citep{pasto08b,smi12,pasto15a}.
Occasionally, SNe Ibn may show multiple light curve peaks \citep[e.g.,][]{gor14}, while
in other cases, a remarkably fast, linear post-maximum decline has been observed  \citep{pasto15b}. In one case,
OGLE-2012-SN-006, a very slow late-time photometric evolution allowed us to monitor the SN for a long time interval
\citep{pasto15c}. The variety in their observables, however, is not surprising, as the evolution of SNe Ibn strongly depends on the 
final configuration of the progenitor star, along with the chemical composition and the geometric properties of the CSM. 
A systematic study of the observed properties of SNe Ibn will be presented in a forthcoming paper. %\citet{pasto15d}.

Most SNe Ibn have been observed in star-forming environments \citep{pasto15b,taddia15}, favouring the association of this SN type
with a massive stars. This association was challenged by the discovery of PS1-12sk in the outskirts of the elliptical galaxy CGCG 208–042
 \citep{san13}. The Type Ibn SN we discuss here,
ASASSN-15ed, appears to robustly connect SNe Ibn with more classical SNe Ib/c (and, hence,
to massive stripped-envelope progenitor stars).

ASASSN-15ed was discovered on 2015 March 1.59 UT by the All Sky Automated Survey for SuperNovae 
\citep[ASAS-SN or ``Assassin'',][]{sha14}, using the four 14-cm ``Brutus'' robotic 
telescopes located in the Haleakala station (Hawaii, USA) of the Las Cumbres Observatory Global Telescope Network
(LCOGT)\footnote{{\it http://lcogt.net/}}. The coordinates of the objects are: R.A. = 16:48:25.16 Dec = +50:59:30.7.
The discovery magnitude reported by \citet{fer15} is 17.1 mag, while there was no detection to a limit of 17.6 mag
 in an image taken on 2015 February 25.57 UT. Although this detection limit is not sufficiently deep to firmly constrain the explosion 
epoch, the available early photometry allows us to state that the object was still increasing in luminosity at the time of discovery.

This paper is organized as follows. In Section \ref{host}, we provide some basic information on the host galaxy,
and estimate the distance and the extinction along the line of sight. In Section \ref{photometry}, the light curve of
 ASASSN-15ed is presented, and the photometric parameters are compared with those of other stripped-envelope SNe. 
In Section \ref{spectroscopy}, the spectra of ASASSN-15ed are shown, and their properties are analysed 
in the context of SNe Ibn. Finally, a discussion and a summary follow in Section \ref{ds}.

\section{Main parameters of MCG +09-27-087} \label{host}

\begin{figure}
\includegraphics[width=8.6cm,angle=0]{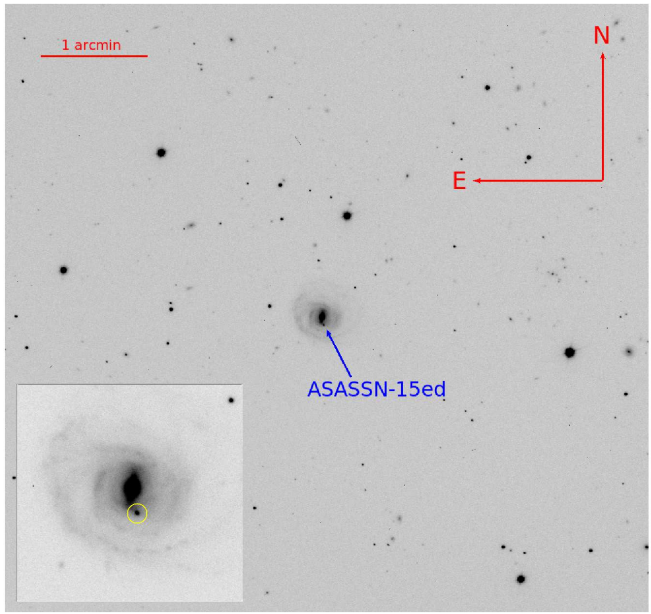} 
% \vspace{302pt}
 \caption{NOT + ALFOSC $r$-band image taken on 2015 April 12, showing the field of ASASSN-15ed, with the host galaxy MCG +09-27-087. 
A blow-up of the galaxy with the SN (marked with a yellow circle) is shown in the bottom-left panel. \label{fig1}}
\end{figure}

Very little information is available on the galaxy hosting ASASSN-15ed,
MCG +09-27-087 (see Figure \ref{fig1}). It is a nearly face-on ring galaxy with total apparent magnitude $g = 15.42 \pm 0.01$
and colours $u - g = 1.79 \pm 0.03$, $g - r = 0.74 \pm 0.01$, $r - i = 0.36 \pm 0.01$ and $i - z = 0.27 \pm 0.01$
\citep[by Sloan Digital Sky Survey Data Release 6 as obtained on 2007 September 17,][]{ade08}.
In the late-time spectra of ASASSN-15ed,  a weak and unresolved H$\alpha$ line in emission 
was detected, from which we infer a redshift z =  0.04866 $\pm$ 0.00017 for the host galaxy.
Adopting a standard cosmology with H$_0$ = 73 km s$^{-1}$ Mpc$^{-1}$, $\Omega_M$ = 0.27, $\Omega_\Lambda$ = 0.73,
we obtain a luminosity distance of 207.5 $\pm$ 13.7 Mpc, which provides a distance modulus
$\mu$ = 36.59 $\pm$ 0.14 mag. 
%mu = 36.585 $\pm$ 0.143

The line-of-sight extinction of ASASSN-15ed is largely dominated by the host galaxy component.
This can be verified by inspecting the higher resolution and best signal-to-noise (S/N) spectra (see Section \ref{spectroscopy}). 
We can safely detect the Na ID feature at the redshift of MCG +09-27-087 in our spectra, with an
equivalent width of 0.8 $\pm$ 0.1 \AA. Following \cite{tur03}, we obtain a colour excess E($B-V$) = 0.12 mag due to the
host galaxy extinction. The same number is obtained adopting the empirical law from \citet{poz12}.
 From our spectra, there is no evidence for the presence of the narrow interstellar Na~I doublet (Na~ID) at z = 0,
therefore we will adopt E($B-V$) = 0.02 mag for the Galactic reddening contribution \citep{sch11}, assuming a reddening
law with R$_V$ = 3.1 \citep{car89}.
Hence, hereafter, we will assume E($B-V$) = 0.14 $\pm$ 0.04 mag as our best estimate for the total colour excess. 

%A_V(MW) = 0.068; A_V(Host) = 0.372; A_V(tot) = 0.440 $\pm$ 0.124
%A_r(MW) = 0.057; A_r(Host) = 0.312; A_r(tot) = 0.369 $\pm$ 0.104
 
\begin{table*}
 \centering
 \begin{minipage}{165mm}
  \caption{Photometric data for ASASSN-15ed. Early 3-sigma limits  from ASAS-SN observations are also reported. \label{photo_tab}}
  \begin{tabular}{cccccccccc}
  \hline
Date        &  JD &  $B $  &   $V$    &   $u$ &  $g$ & $r$ &  $i$     &  $z$   & Instrument  \\
 \hline
20150220 & 2457074.08 &     --     &   $>$18.26   &     --     &       --     &       --     &       --     &       --      &   ASASSN \\
20150225 & 2457079.07 &     --     &   $>$18.11   &     --     &       --     &       --     &       --     &       --      &   ASASSN \\
20150301 & 2457083.09 &     --     &   17.39 0.15 &     --     &       --     &       --     &       --     &       --      &   ASASSN \\
20150302 & 2457083.72 &     --     &       --     &     --     &       --     &   17.21 0.24 &       --     &       --      &   OIM$^1$ \\
20150306 & 2457087.92 &     --     &   17.11 0.26 &     --     &       --     &   16.92 0.25 &       --     &       --      &   NMSO$^2$  \\
20150309 & 2457090.97 &     --     &   17.32 0.24 &     --     &       --     &   17.13 0.22 &       --     &       --      &   NMSO$^2$ \\
20150311 & 2457092.91 &     --     &   17.50 0.23 &     --     &       --     &   17.29 0.25 &       --     &       --      &   NMSO$^2$ \\
20150312 & 2457093.59 &     --     &       --     &     --     &   17.58 0.05 &   17.40 0.10 &       --     &       --      &   AFOSC \\
20150313 & 2457094.61 &     --     &       --     & 18.33 0.08 &   17.73 0.05 &   17.62 0.03 &   17.58 0.04 &   17.69 0.04  &   IO:O \\
20150313 & 2457094.67 & 18.08 0.07 &   17.79 0.05 & 18.40 0.04 &   17.76 0.05 &   17.64 0.06 &   17.62 0.09 &   17.72 0.11  &   AFOSC \\
20150314 & 2457095.51 & 18.26 0.15 &       --     & 18.56 0.19 &       --     &       --     &       --     &       --      &   AFOSC \\
20150314 & 2457095.58 & 18.25 0.12 &   17.88 0.12 & 18.54 0.03 &   17.84 0.06 &   17.68 0.16 &   17.70 0.12 &   17.76 0.27  &   AFOSC \\
20150314 & 2457095.91 & 18.28 0.11 &   17.92 0.08 &     --     &   17.90 0.09 &   17.71 0.16 &   17.74 0.19 &       --      &   LCOGT \\
20150315 & 2457096.58 & 18.32 0.13 &       --     & 18.66 0.05 &   17.99 0.04 &   17.75 0.03 &   17.77 0.03 &       --      &   LRS \\
20150315 & 2457096.61 &     --     &       --     & 18.66 0.12 &   18.02 0.04 &   17.75 0.05 &   17.78 0.04 &   17.79 0.06  &   IO:O \\
20150315 & 2457096.93 & 18.40 0.06 &   17.97 0.10 &     --     &   18.09 0.05 &   17.79 0.08 &   17.83 0.13 &       --      &   LCOGT \\
20150316 & 2457097.96 & 18.52 0.09 &   18.11 0.07 &     --     &   18.23 0.07 &   17.96 0.10 &   17.87 0.13 &       --      &   LCOGT \\
20150317 & 2457098.61 &     --     &       --     & 19.17 0.08 &   18.34 0.04 &   18.08 0.03 &   17.96 0.05 &   17.90 0.06  &   IO:O \\
20150319 & 2457100.59 & 18.93 0.12 &   18.35 0.13 & 19.54 0.08 &   18.64 0.14 &   18.29 0.17 &   18.21 0.18 &   18.20 0.23  &   AFOSC \\
20150322 & 2457103.87 & 19.49 0.13 &   18.75 0.12 &     --     &   18.97 0.11 &   18.64 0.12 &   18.60 0.15 &       --      &   LCOGT \\
20150323 & 2457105.46 & 19.73 0.14 &   18.96 0.19 & 20.39 0.41 &   19.18 0.19 &   18.82 0.26 &   18.87 0.28 &   18.69 0.23  &   AFOSC \\
20150324 & 2457105.82 & 19.82 0.17 &   19.00 0.14 &     --     &   19.21 0.12 &   18.87 0.16 &   18.93 0.25 &       --      &   LCOGT \\
20150325 & 2457106.81 & 19.99 0.19 &   19.12 0.19 &     --     &   19.37 0.14 &   19.05 0.19 &   19.07 0.18 &       --      &   LCOGT \\
20150328 & 2457109.64 &     --     &       --     &     --     &       --     &   19.20 0.12 &       --     &       --      &   OSIRIS \\
20150328 & 2457109.74 & 20.23 0.04 &   19.32 0.04 & 20.88 0.05 &       --     &   19.21 0.03 &   19.21 0.04 &   18.93 0.05  &   ALFOSC \\
20150328 & 2457109.76 &     --     &       --     & 20.92 0.33 &   19.63 0.10 &   19.22 0.08 &   19.24 0.10 &   18.92 0.16  &   IO:O \\
20150330 & 2457111.54 &     --     &       --     & $>$18.73 0 &   19.77 0.46 &   19.47 0.35 &       --     &   19.03 0.45  &   IO:O \\
20150331 & 2457112.57 & 20.75 0.22 &   19.61 0.36 &     --     &   19.86 0.17 &   19.53 0.14 &   19.44 0.18 &   19.15 0.27  &   AFOSC \\
20150401 & 2457113.54 &     --     &       --     & 21.56 0.48 &   19.97 0.12 &   19.61 0.11 &   19.52 0.09 &   19.24 0.12  &   IO:O \\
20150405 & 2457117.57 &     --     &       --     &     --     &   20.29 0.12 &   19.82 0.10 &   19.77 0.11 &   19.50 0.11  &   IO:O \\
20150405 & 2457117.71 &     --     &       --     &     --     &       --     &   19.84 0.18 &       --     &       --      &   OSIRIS \\
20150408 & 2457120.70 &     --     &       --     &     --     &   20.52 0.06 &   20.00 0.14 &   20.05 0.24 &   19.72 0.21  &   IO:O \\
20150410 & 2457122.63 &     --     &       --     &     --     &   20.70 0.17 &   20.13 0.09 &   20.21 0.14 &   19.81 0.10  &   IO:O \\
20150412 & 2457124.57 &     --     &       --     &     --     &   20.85 0.17 &   20.28 0.12 &   20.40 0.18 &   20.00 0.14  &   IO:O \\
20150412 & 2457124.63 &     --     &       --     &     --     &   20.87 0.09 &   20.31 0.07 &   20.43 0.06 &   20.02 0.10  &   ALFOSC \\
20150415 & 2457127.59 &     --     &       --     &     --     &   21.21 0.07 &   20.76 0.06 &   20.71 0.06 &   20.08 0.11  &   IO:O \\
20150416 & 2457128.61 & 22.36 0.68 &   21.16 0.54 &     --     &       --     &       --     &       --     &       --      &   AFOSC \\
20150418 & 2457131.50 &     --     &       --     &     --     &   21.40 0.08 &   20.96 0.07 &   21.01 0.10 &   20.36 0.11  &   IO:O \\
20150423 & 2457135.62 &     --     &       --     &     --     &   21.84 0.11 &   21.26 0.10 &   21.19 0.11 &   20.48 0.12  &   IO:O \\
20150425 & 2457137.57 &     --     &       --     &     --     &   21.94 0.09 &   21.43 0.10 &   21.34 0.12 &   20.65 0.14  &   IO:O \\
20150426 & 2457139.50 &     --     &       --     &     --     &   22.10 0.15 &   21.78 0.17 &   21.60 0.16 &   20.96 0.23  &   IO:O \\
20150428 & 2457140.62 &     --     &       --     &     --     &   22.30 0.11 &   22.19 0.14 &   21.85 0.20 &   $>$20.00    &   OSIRIS \\
20150428 & 2457140.69 &     --     &   22.36 0.34 &     --     &   22.35 0.21 &       --     &       --     &       --      &   LRS \\
20150429 & 2457142.48 &     --     &       --     &     --     &   $>$22.07   &   22.70 0.36 &   22.23 0.27 &   21.62 0.19  &   IO:O \\
20150509 & 2457151.62 &     --     &    --        &     --     &   $>$23.54   &  $>$23.23    &   $>$23.07   &       --      &   ALFOSC \\
20150511 & 2457153.64 &     --     &    --        &     --     &   $>$23.62   &  $>$23.52    &   $>$23.08   &   $>$22.40    &   OSIRIS \\ 
 \hline
\end{tabular}

$^1$ 20-cm Celestron C8 telescope with Mammut L429 Camera + Sony ICX420 CCD; Observatory Inmaculada del Molino (OIM), Osu\~na (Spain);\\
$^2$ 51-cm RCOS telescope with SBIG STXL-6303 camera, New Mexico Skies Observatory (NMSO), Mayhill (New Mexico, USA).
\end{minipage}
\end{table*}

As the total $B$-band magnitude of MCG +09-27-087 is $B$ = 16.03, we obtain an absolute
magnitude of $M_B$ = $-$20.67, corrected for Galactic and internal extinction following the 
prescriptions of the Hyperleda database\footnote{{\it http://leda.univ-lyon1.fr/}; \citep{mak14}.}. Following \citet{tre04}, we estimate an integrated 
oxygen abundance of
12 + log(O/H) =  9.08 (dex). Accounting for the position of the SN in the host galaxy (1.0 arcsec west and 4.9 arcsec south
from the galaxy core) and following \citet{pasto15b}, we compute the R$_{SN}/$R$_{25}$ ratio\footnote{R$_{SN}$ is the deprojected position of the SN and
R$_{25}$ is the isophotal radius for the $B$-band surface brightness of 25 mag arcsec$^{−2}$.} ($\sim$ 0.35).  This is used 
to estimate an oxygen abundance of 12 + log(O/H) =  8.60 (dex) at the position of ASASSN-15ed, after adopting the radial metallicity gradient from
\citet{pil04}. This is close to the average values obtained by \citet{taddia15}
and \citet{pasto15b} for the galaxies hosting SNe Ibn (about 8.4-8.6 dex).

\section[]{Photometry} \label{photometry}

As mentioned before, ASASSN-15ed was discovered by the ASAS-SN collaboration, hence we included in our data set the discovery image and the
two closest non-detections. These data were obtained with the quadruple 14-cm ``Brutus'' Telescope.
The early-time photometric evolution of ASASSN-15ed was recovered by inspecting unfiltered images collected
with amateur instruments (see Table \ref{photo_tab}, for more information). 
Our multi-filter follow-up campaign started on 2015 March 12, when the object was already past-maximum, and lasted over two months. 
It was performed using several telescopes available to the collaboration, including the 1.82-m Copernico Telescope at Mt. Ekar (Asiago, Italy), 
equipped with AFOSC, the LCOGT 1.0-m telescope at McDonald Observatory (Texas, USA) equipped with an SBIG camera,
the 10.4-m Gran Telescopio Canarias (GTC) equipped with OSIRIS,  the 3.58-m Telescopio Nazionale Galileo (TNG) equipped with DOLORES (LRS), the 2.5-m Nordic Optical Telescope (NOT) 
with ALFOSC and the 2.0-m Liverpool Telescope (LT) with the IO:O camera, all sited at La Palma (Canary Islands, Spain).

The imaging frames were first pre-reduced (i.e. overscan, bias and flat-field corrected, and trimmed in order to remove the unexposed regions of the image)
using standard routines in   \textsc{IRAF}\footnote{IRAF is distributed by the National Optical Astronomy Observatory,
which is operated by the Association of Universities for Research in Astronomy (AURA) under cooperative agreement with the National Science Foundation.}. 
The subsequent steps of data reduction were performed using a dedicated \textsc{PYTHON}-based pipeline, \textsc{SNOoPY} \citep{ec14},
which is a collection of \textsc{PYRAF} tasks and other public tools (including \textsc{DAOPHOT}, \textsc{SEXTRACTOR}, \textsc{HOTPANTS}) to astrometrically register the images, and extract and measure the magnitudes of
the stellar sources detected in the images using PSF-fitting techniques. The SN magnitude was measured after subtracting the host galaxy background using a
low-order polynomial fit.

As the SN field was observed by Sloan Digital Sky Survey (SDSS), zeropoints and colour terms for individual nights  for Sloan ($u$, $g$, $r$, $i$, and $z$) filters were obtained using the catalogued SDSS 
magnitudes for a large number of field stars. The reference $B$ and $V$ bands magnitudes of the field stars were derived by converting their
Sloan magnitudes to Johnson-Bessell magnitudes, following the prescriptions of \citet{cho08}.

Early unfiltered data collected from  amateur astronomers were scaled to Johnson $V$ and/or Sloan $r$ magnitudes according to the quantum 
efficiency curves of the different CCDs used for the observations. The final, calibrated SN magnitudes and their uncertainties are reported in Table \ref{photo_tab}, 
along with the basic information on the 
instruments used for the photometric follow-up campaign.

Errors were estimates using artificial star experiments. Fake stars with magnitudes similar to that
of the SN are placed in positions close to the SN. The rms of the recovered instrumental point spread function (PSF) magnitudes 
accounts for the background fitting uncertainty. This value is combined (in quadrature) with the PSF-fit
error returned by \textsc{DAOPHOT}, providing the total instrumental magnitude error. 
Finally, we propagated errors from the photometric calibration.

\subsection[]{Light curves} \label{lc}

\begin{figure*}
\includegraphics[width=13.5cm]{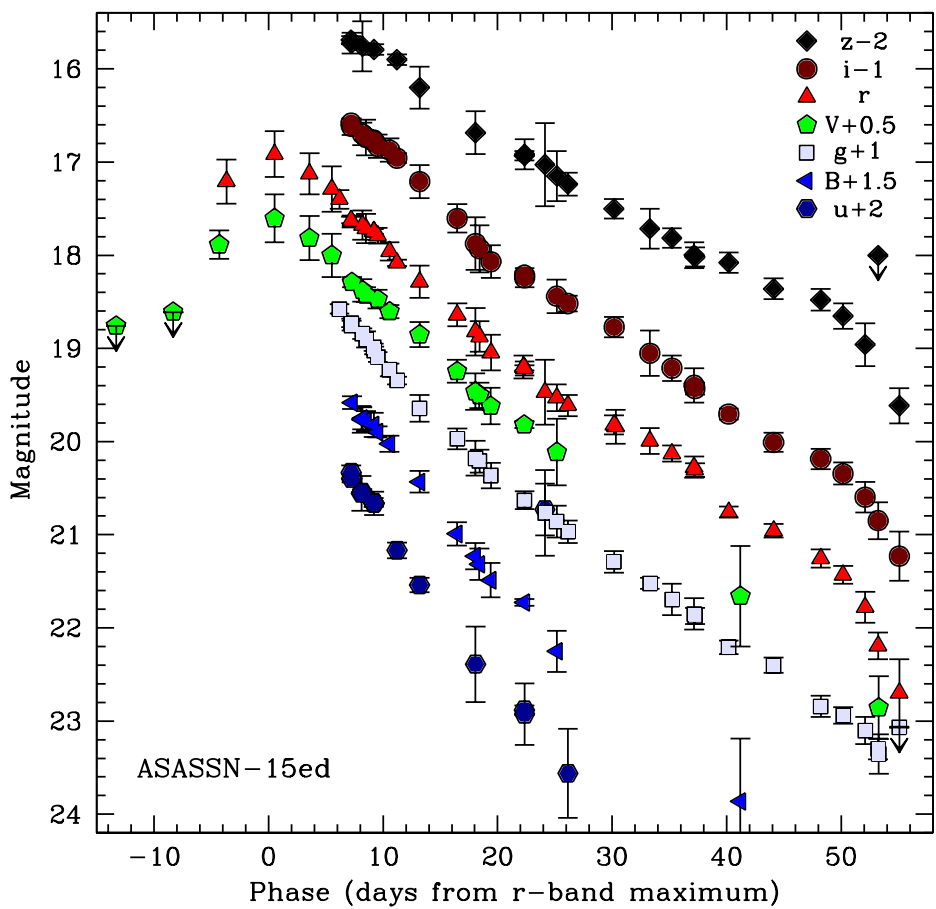} 
%\vspace{302pt}
 \caption{Multi-band light curves of ASASSN-15ed. Phase 0 is coincident with the epoch of the $r$-band maximum light, JD = 2457087.4 $\pm$ 0.6.\label{fig2}}
\end{figure*}

Johnson $BV$ and Sloan $ugriz$ light curves of ASASSN-15ed are shown in Figure \ref{fig2}. The pre-maximum 
evolution is not well constrained, and with the available data we can only constrain the rise time to be $>$ 4.3 d.
Occasionally, SNe Ibn reach the maximum light in less than 4~d \citep[e.g. SNe~1999cq and LSQ12btw,][]{mat00,pasto15b}. 
Using a low-order polynomial fit, we estimate that the $r$-band maximum light occurred on 2012 March 5.9 UT (JD = 2457087.4 $\pm$ 0.6).
This allows us to estimate the $r$-band peak magnitude to be $r = 16.91 \pm 0.10$. Accounting for the reddening and distance estimate provided in Section \ref{host},
we estimate an absolute magnitude at maximum of $M_r = -20.04 \pm 0.20$. As a consequence, ASASSN-15ed is
one of the highest luminosity events of its class \citep{pasto15e}.

\begin{table}
\begin{center}
%\scriptsize
\caption{Light curve parameters for ASASSN-15ed. \label{lcpar} }
\begin{tabular}{ccccc}
\hline \hline
Filter   & peak magnitude & $\gamma_{0-25}^\ddag$ & $\gamma_{25-50}^\ddag$ & $\gamma_{>50}^\ddag$\\ \hline
$B$ & --             & 14.96 $\pm$ 0.26 & 10.88 $\pm$ 1.07 & -- \\
$V$ & 17.10$\pm$0.11 & 10.64 $\pm$ 0.18 & 9.51 $\pm$ 0.24 & -- \\
$u$ & --             & 16.95 $\pm$ 0.38 &  -- & -- \\
$g$ & --             & 12.08 $\pm$ 0.26 & 8.58 $\pm$ 0.14 & -- \\
$r$ & 16.91$\pm$0.10 & 11.37 $\pm$ 0.23 & 7.70 $\pm$ 0.26 & 26.32 $\pm$ 2.47 \\
$i$ & --             & 11.43 $\pm$ 0.28 & 7.81 $\pm$ 0.18 & 18.24 $\pm$ 1.57 \\
$z$ & --             & 8.27 $\pm$ 0.31 & 5.86 $\pm$ 0.23 & 16.81 $\pm$ 2.19 \\
 \hline
\end{tabular}
\begin{flushleft}

$^\ddag$ in mag/100$^d$ units
\end{flushleft}
\end{center}
\end{table}

Following \citet{pasto15d}, we estimated the light curve decline rates in the different bands for ASASSN-15ed, with the results listed in Table \ref{lcpar}.
ASASSN-15ed has a rapid decline during the first 3-4 weeks in all bands, with the blue band light curves declining faster than those in the red bands 
(the $u$-band decline rate during the first 25 d after maximum -- $\gamma_{0-25}(u) \approx$ 0.17 mag d$^{-1}$ -- is twice as fast as the $z$-band, $\gamma_{0-25}(z) \approx$ 0.08 mag d$^{-1}$). 
From 25 d to 50 d, the light curves flatten. At these phases, the decline rate is 
0.08-0.09 mag d$^{-1}$ in all bands (the $B$-band light curve seems to decline more rapidly, but
the decay rate is computed with two points only). 
Later on, at phases later than 50 d, the light curve steepens. For these late phases, decline rate measurements were only made for the 
red bands, since now  the SN luminosity declined
below the detection threshold at bluer wavelengths. In the $r$-band, the light curve decline rate at phases above 50 d is extremely fast, 
being $\gamma_{>50}(r) \approx$  0.26 mag d$^{-1}$, 
while it is slightly slower in the $i$ and $z$ bands (0.17-0.18 mag d$^{-1}$).
This remarkable 
weakening in the optical luminosity at late phases is not unprecedented in SNe Ibn \citep{matt08,smi08} and is likely a signature of dust formation.

\subsection[]{Photometric Comparison with Other He-rich SNe} \label{lc_cmp}

In order to better understand ASASSN-15ed in the context of He-rich core-collapse SNe,
we calculated a pseudo-bolometric light curve using the collected optical photometry.
The pseudo-bolometric luminosity was computed at the epochs when $r$-band observations were available. When
 photometric points were not available in a given filter at some epochs, the contribution
of the missing observations was estimated through an interpolation of the magnitudes
at adjacent epochs.
As pre-maximum photometry was available only in the $r$ and $V$ bands, the early flux contributions
from the other bands was computed by extrapolating the colour information using
the available multi-band photometry and early-time data of SN~2010al.

\begin{figure*}
\includegraphics[width=16cm,angle=0]{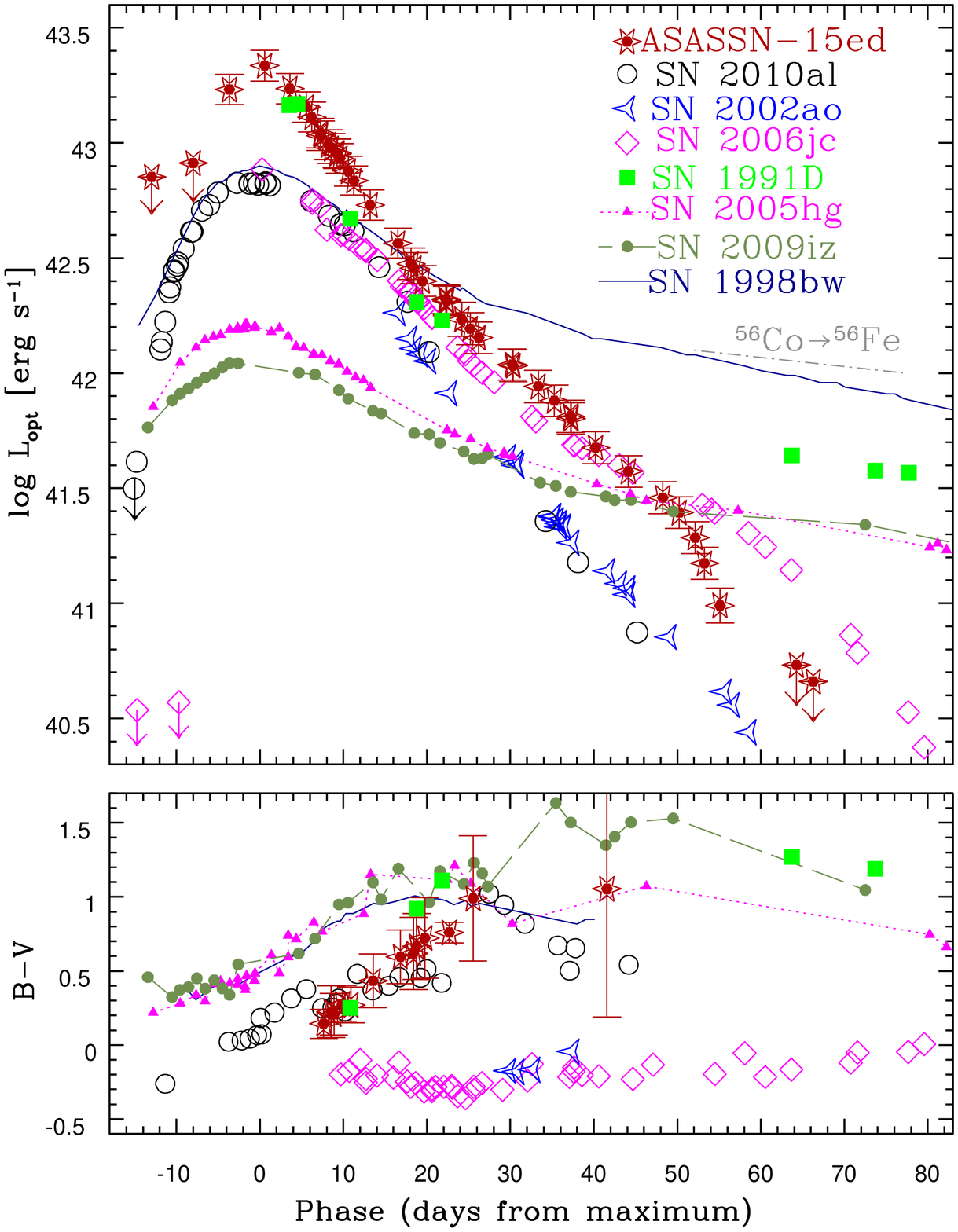} 
% \vspace{302pt}
 \caption{Top - Pseudo-bolometric light curve of ASASSN-15ed ($uBVgriz$ bands), compared with those of a sample of reference stripped-envelope SNe. As comparison objects, we show
SNe Ibn 2010al \protect\citep[$UBVRI$ bands; $\mu$ = 34.27 mag, E($B-V$) = 0.06 mag;][]{pasto15a}, 2002ao \protect\citep[$BVRI$ bands; $\mu$ = 31.73 mag, E($B-V$) = 0.25 mag;]{pasto08a} and 
2006jc  \protect\citep[$UBVRI$ bands; $\mu$ = 32.01 mag, E($B-V$) = 0.04 mag;][]{pasto07,fol07,pasto08a,anu09};
the peculiar SN Ib 1991D \protect\citep[$BVR$ bands; $\mu$ = 36.24 mag, E($B-V$) = 0.05 mag;][]{ben02}; the normal SNe Ib 2005hg ($uBVri$ bands; $\mu$ = 34.80 mag, E($B-V$) = 0.09 mag) and 
2009iz \protect\citep[$uBVri$ bands; $\mu$ = 33.92 mag, E($B-V$) = 0.07 mag;][]{bia14}; the luminous, broad-lined
Type Ic SN 1998bw \citep[$UBVRI$ bands; $\mu$ = 32.74 mag, E($B-V$) = 0.06 mag;][and references therein]{pat01}.  Bottom - Comparison of the $B-V$ colour evolution for the same set of SNe.
\label{fig_bolo}}
\end{figure*}

\begin{table*}
 \centering
 \begin{minipage}{140mm}
  \caption{Log of spectroscopic observations of ASASSN-15ed. \label{spec_tab}}
  \begin{tabular}{cccccccccc}
  \hline
Date        &  JD &  Phase (d)$^\ddag$ &   Instrumental configuration    &  Exposure time (s) &  Range (\AA) & Resolution  (\AA) \\
 \hline
2015/03/11 & 2457092.97 & +5.6 & 2.4-m Hiltner + OSMOS + VPH & 3 $\times$ 1200 & 3980--6860 & 3.4 \\
2015/03/12 & 2457093.61 & +6.2 & 1.82-m Copernico + AFOSC + gm4 & 2400 & 3360--8200 & 14 \\
2015/03/13 & 2457094.65 & +7.3 & 1.82-m Copernico + AFOSC + gm4 & 2$\times$2100 & 3470--8190 & 14 \\
2015/03/14 & 2457095.64 & +8.2 & 1.82-m Copernico + AFOSC + gm4 & 3$\times$ 1800 & 3500--8190 & 14 \\ %clouds
2015/03/15 & 2457096.58 & +9.2  & 3.58-m TNG + LRS + LRB & 2400 & 3250--8160 & 10 \\
2015/03/16 & 2457097.99 & +10.6 & 1.5-m FLWO + FAST + 300 gpm & 1800 & 3470--7410 & 5.7 \\ 
2015/03/17 & 2457098.96 & +11.6 & 1.5-m FLWO + FAST + 300 gpm & 3 $\times$ 1200  & 3470--7410 & 5.7 \\ 
2015/03/19 & 2457100.56 & +13.2 & 1.82-m Copernico + AFOSC + gm4 & 3600 & 3450--8190 & 14 \\
2015/03/19 & 2457101.02 & +13.6 & 2.0-m FTN + FLOYDS  & 3600 & 3200-10000 & 13 \\
2015/03/23 & 2457105.00 & +17.6 & 2.0-m FTN + FLOYDS  & 3600 & 3200-10000 & 13 \\
2015/03/24 & 2457105.51 & +18.1 & 1.82-m Copernico + AFOSC + gm4 & 3600 & 3550--8190 & 14 \\
2015/03/26 & 2457107.57 & +20.2 & 3.58-m TNG + LRS + LRR & 3600 & 5100-10550 & 14 \\
2015/03/28 & 2457109.67 & +22.3 & 10.4-m GTC + OSIRIS + R1000B & 2 $\times$ 1800 & 3630--7880 & 7.0 \\
2015/04/05 & 2457117.73 & +30.4 & 10.4-m GTC + OSIRIS + R1000R & 6 $\times$ 600 &  5100-10400 & 7.7 \\
2015/04/28 & 2457140.57 & +53.2 & 10.4-m GTC + OSIRIS + R1000R & 1800 & 5100-10400 & 7.7 \\
 \hline
\end{tabular}
$^\ddag$  The phases are with respect to the $r$-band maximum light (JD = 2457087.4).
\end{minipage}
\end{table*}

The fluxes were corrected for the extinction reported in Section \ref{host}, and then integrated using 
the trapezoidal rule. The observed fluxes were finally converted to luminosity using the host galaxy distance 
adopted in this paper. The resulting pseudo-bolometric light curve of ASASSN-15ed is shown in Figure \ref{fig_bolo} (top panel),
and the data are listed in Table \ref{lcbol} in the Appendix.
It may significantly differ from a ``true'' bolometric light curve, particularly
at early phases when the contribution from the unobserved UV bands can be dominant, as seen in other
SNe Ibn \citep[e.g. SN 2010al, see][]{pasto15a}, and at late phases when the contribution of the IR flux
might be large in the case of dust condensation  \citep[like in SN 20006jc,][]{matt08,smi08}.

We compared the quasi-bolometric light curve of ASASSN-15ed to those for a wide sample of stripped envelope SNe, including three
SNe Ibn (2002ao, 2010al and 2006jc), two classical SNe Ib (2005hg and 2009iz), the luminous and fast-evolving SN Ib 1991D
and the prototypical hypernova (Type Ic) SN 1998bw. References for the data are reported in the caption of Figure \ref{fig_bolo}.
We note that the peak luminosity of ASASSN-15ed largely exceeds that of other stripped-envelope SNe shown in the figure, with the possible exception
of the puzzling SN 1991D  \citep{ben02}. SN~1991D was spectroscopically classified as a relatively normal SN Ib, but
 was remarkably luminous and had an unusually fast-evolving light curve. The pseudo-bolometric peak luminosity of SN 1991D has to be regarded 
as a lower limit, since the epoch of maximum light is
not well constrained, and the quasi-bolometric light curve has been computed  accounting for the contribution
of solely three optical bands ($B, V$ and $R$), as the SN was not imaged with other filters.

Another interesting property of the pseudo-bolometric light curves of ASASSN-15ed and other SNe Ibn shown in Figure \ref{fig_bolo}
is the fast and almost linear post-peak decline
up to phase $\sim$ 50-60 d, without flattening on to the $^{56}$Co tail. Again, the steep luminosity decline is similar to that observed 
in SN 1991D. However, in SN 1991D, we note a clear flattening $\sim$ 2 months after peak,
which is not observed in ASASSN-15ed or the other SNe Ibn discussed here.\footnote{A late-time light curve flattening was observed only
in the Type Ibn SN OGLE-2012-SN-006, but was attributed to strong ejecta-CSM interaction \protect\citep{pasto15c}.} 
Instead, ASASSN-15ed began an even steeper decline starting  $\sim$ 2 months
after maximum, and confirmed by detection limits at phases $>$ 64 d.

\begin{figure*}
\includegraphics[width=16cm,angle=0]{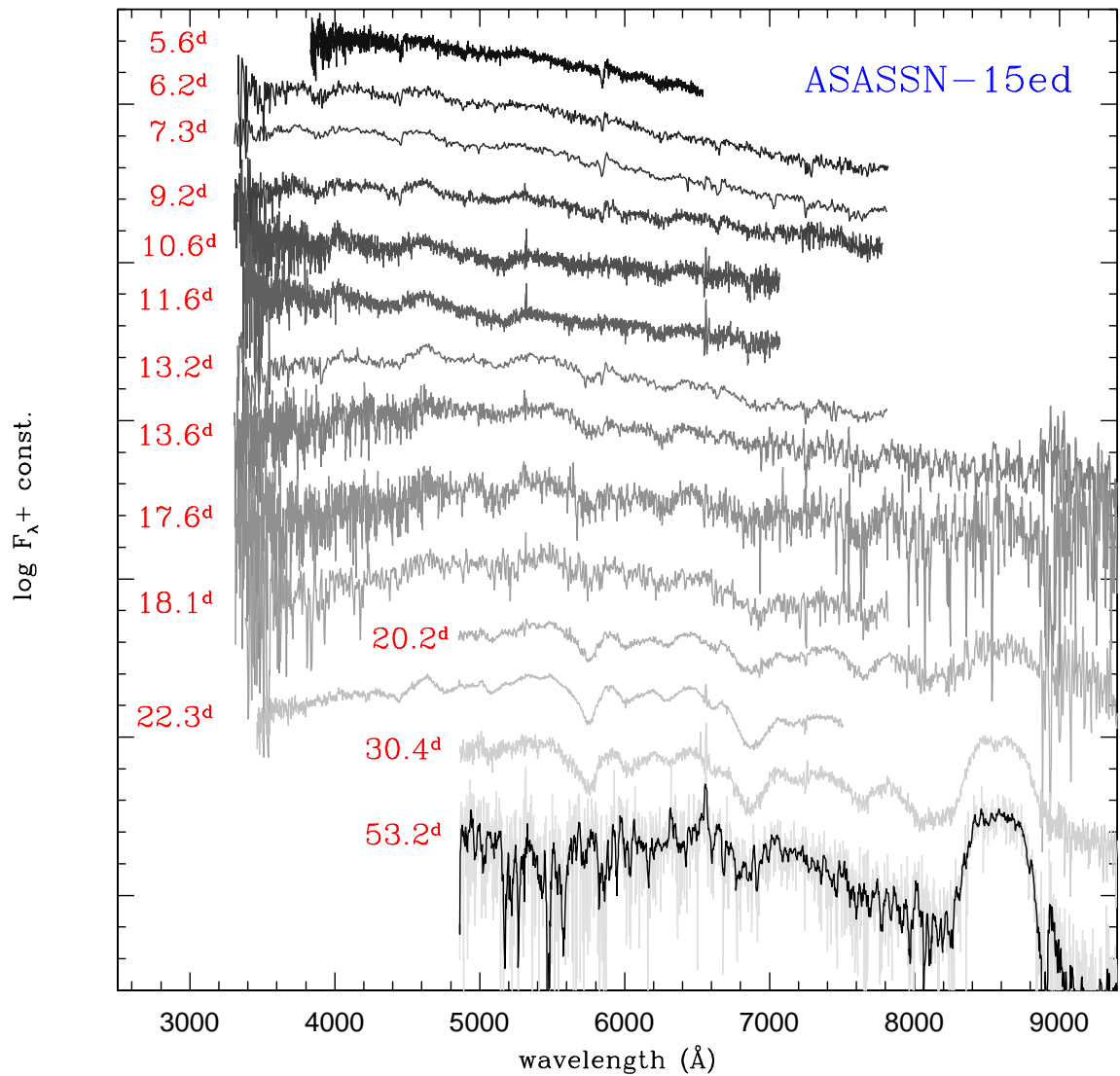} 
% \vspace{302pt}
 \caption{Spectral sequence of ASASSN-15ed. The spectra have been redshift-corrected, but no reddening correction has been applied.\label{fig3}.
The phases indicated on left are with respect to the epoch of the $r$-band maximum.
The spectrum at phase +53.2 d has been smoothed using a boxcar with size 9 pixels. \label{seqspec}}
\end{figure*}

Although the lack of a clear light curve flattening  on to the $^{56}$Co decay rate would argue against a significant 
amount of $^{56}$Ni synthesized in the explosion, we note that the quasi-bolometric light curve of  ASASSN-15ed is very similar
to that of SN 2006jc if considering only the optical bands. In this case, however, the late-time
optical deficit was balanced by a strong IR excess \citep[e.g.][]{dica08,matt08,smi08}, which may be consequence of dust formation
in a cool dense shell. 
Unfortunately, ASASSN-15ed was not observed in the NIR bands, so we cannot determine whether its weak late-time
luminosity is due to a very small synthesized  $^{56}$Ni  mass, or to dust condensation.

In Figure \ref{fig_bolo} (bottom), we also show the $B-V$ colour evolution of ASASSN-15ed along with those of the same SNe as mentioned above.
Typical non-interacting stripped envelope SNe in our sample have a rather similar colour evolution,
 with $B-V$ increasing from $\sim$ 0.3 mag to nearly 1 mag during the rise to maximum and the subsequent $\sim$ 25 d.
At later phases, their $B-V$ colour behaviour is more heterogeneous, with $B-V$ colours varying in the interval 0.8$-$1.5 mag.

The colour evolution of SNe Ibn is somewhat different. 
Although a few of them (e.g. SNe~2002ao and 2006jc) have a very flat colour evolution, others (including ASASSN-15ed and SN~2010al) $-$ while being very blue at early times $-$
become redder more rapidly with time, placing their behaviour between those of SNe Ib/c and the most extreme
SNe Ibn. We remark, however, that the late-time $B-V$ colour of SN~2006jc was affected by the major weakening of the optical flux
at blue wavelengths due to dust formation, which largely affected the $B$ and $V$ magnitude estimates.
Interestingly, ASASSN-15ed and SN~2010al are the two examples of SNe Ibn which spectroscopically develop
broad P-Cygni line profiles at later phases in their spectra (cf. Section \ref{spc_cfr}). The spectroscopic metamorphosis
and the $B-V$ colour transition from blue colours typical of SNe Ibn to red colours expected in non-interacting stripped-envelope 
SNe happen nearly at the same time (Figure \ref{fig_bolo}, bottom). All of this suggests that these two SNe are transitional objects
between SNe Ib/c and more typical SNe Ibn.

\section[]{Spectroscopy} \label{spectroscopy}

We obtained spectroscopic observations using OSMOS on the 2.4-m Hiltner Telescope of the MDM Observatory
on Kitt Peak (Arizona, USA), FAST on the 1.5-m Tillinghast Telescope of Fred L. Whipple Observatory at Mt. Hopkins (Arizona, USA),
FLOYDS on the 2.0-m Faulkes Telescope North (FTN) of the LCOGT network at Haleakala Observatory (Hawaii, USA), AFOSC on the 1.82-m Copernico Telescope of 
Mt. Ekar (Asiago, Italy), LRS on the 3.58-m TNG and OSIRIS on the the 10.4-m GTC, both operated at La Palma (Spain).

Spectroscopic observations of ASASSN-15ed were carried out between 2015 March 11 (5.6 d after maximum) and April 28 (phase $\sim$ 53 d), and cover the post-peak SN luminosity decline. 
The data were processed using standard IRAF tasks, with the inclusion of bias, flat and overscan corrections. Optimized one-dimensional spectra were
extracted from the two-dimensional frames. Wavelength calibration was performed using reference arc spectra obtained with the same instrumental set-up.  
The accuracy of the wavelength calibration was verified measuring the  position of night sky lines, and $-$ if necessary $-$ a shift in wavelength was applied.
The flux calibration was performed through the spectroscopic observation of at least one flux standard star observed during the same night as the SN.
The  flux calibration of the SN spectrum was then checked against broad-band photometry obtained on the nearest night (ideally, the same night)
and, when necessary, a scaling factor was applied. The spectra of standard stars were also used to correct the SN spectra from the O$_2$ and OH telluric bands.

Information on the final spectra of ASASSN-15ed, and the instrumental configurations is listed in Table \ref{spec_tab}, and the spectra are shown in Figure \ref{seqspec}.

\subsection[]{Spectral Evolution} \label{spc_ev}

\begin{figure*}
{\includegraphics[width=11.25cm,angle=270]{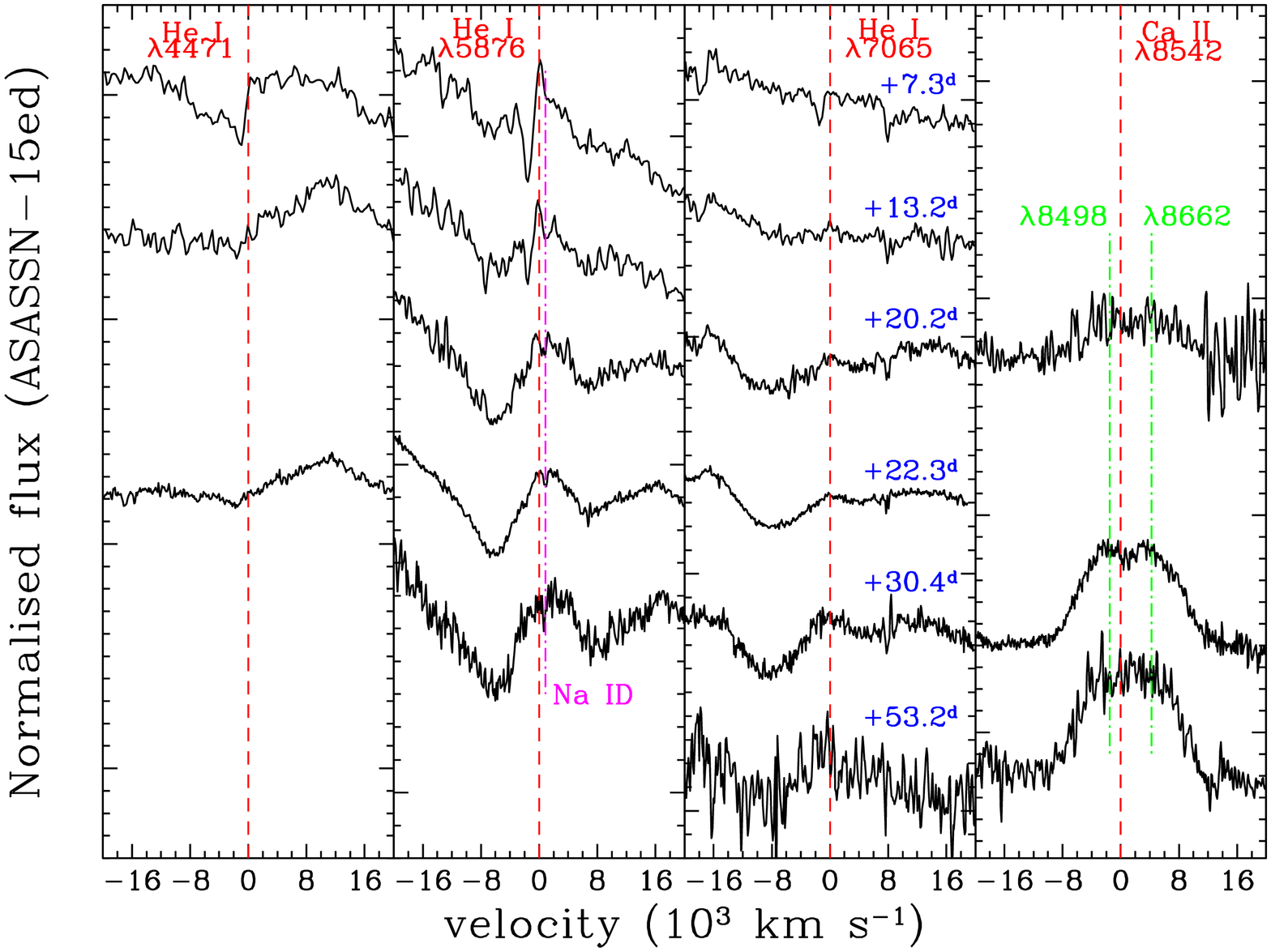} 
\includegraphics[width=11.25cm,angle=270]{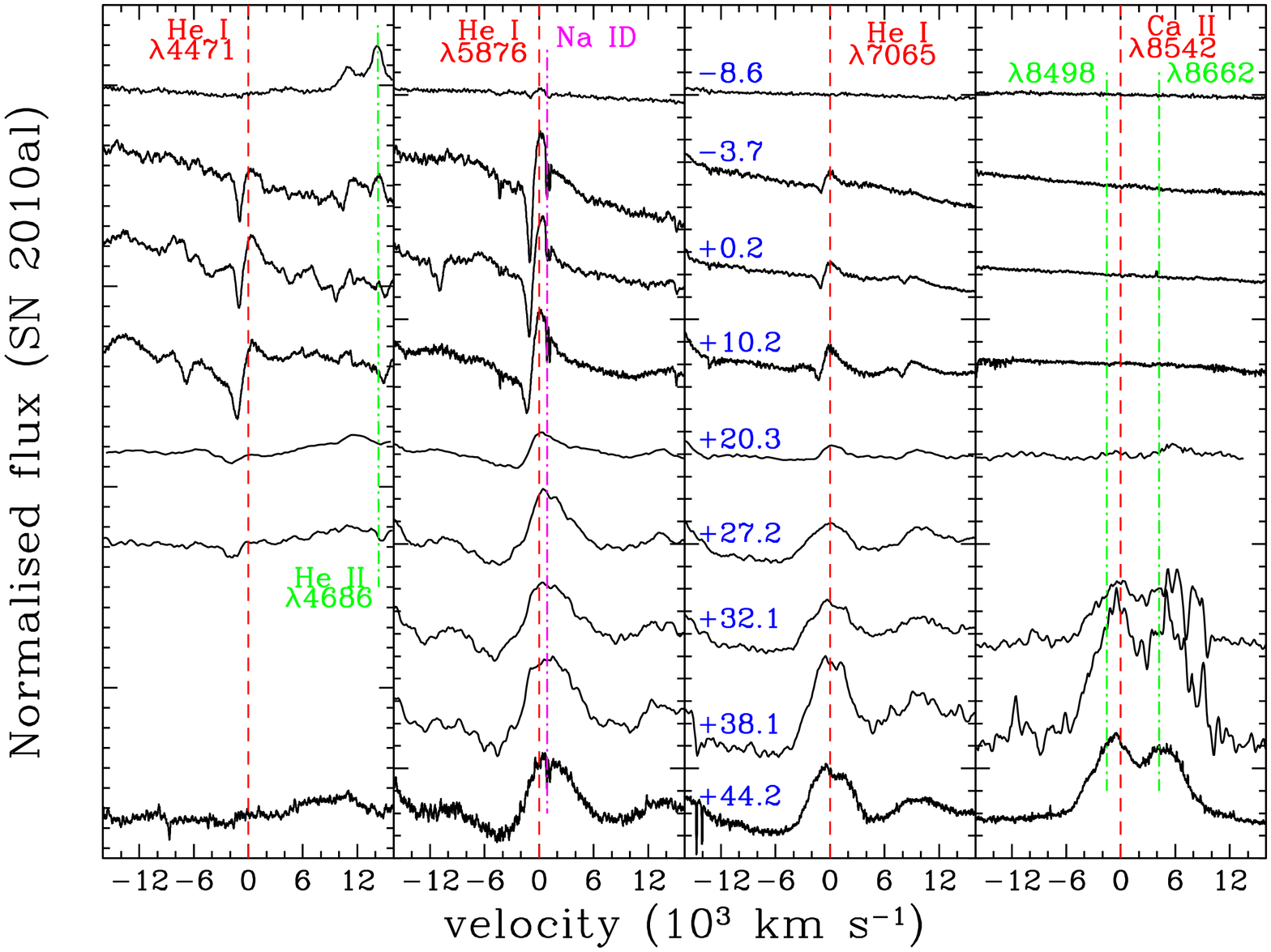}} 
% \vspace{302pt}
 \caption{Evolution of He I $\lambda$4471, He I $\lambda$5876, He I $\lambda$7065, and the NIR Ca II triplet
in selected spectra of ASASSN-15ed (top) and SN~2010al \protect\citep[bottom,][]{pasto15a}. A separation
constant has been applied to the normalized fluxes for clarity.
The spectra are shown on a velocity scale, with the rest wavelength (zero velocity) being indicated with a red dashed vertical line. Dot-dashed 
magenta and green lines mark the positions of other interesting lines.
 \label{seqlines}}
\end{figure*}

Early-time spectra (phases from +5.6 d to +7.3 d) are characterized by an almost featureless blue continuum,
with a  temperature T$_{bb} \approx $ 11000 K, and only weak narrow P-Cygni lines of He I 
are detected. Such a smooth continuum was observed in other young SNe Ibn \citep[e.g. in SN~2000er;][]{pasto08a} 
and was proposed to originate in a thin cool dense shell \citep{chu09}.
From the position of the minimum of the absorption components, we estimate that the gas
producing the narrow lines is moving at about 1200-1300 km s$^{-1}$. 
We also note that the narrow P-Cygni He~I features become weaker with time, and are only barely detected
in later spectra (up to about 3 weeks after maximum), with velocities that are marginally higher, but always less than 1500 km s$^{-1}$.
The negligible velocity evolution of the narrow components suggests that these features are likely
produced in the unshocked He-rich CSM. We note that He-rich gas moving at 1200-1500 km s$^{-1}$
is compatible with mass-loss from a Wolf-Rayet (WR) star \citep[e.g.,][]{tor86,pri90,ham06}. In particular, these pre-SN
wind velocities are compatible with those observed in winds of WNL stars \citep{cro95a,cro95b,hai14}.

In our intermediate-age spectra (from 9.2 d to 11.6 d), very broad and shallow undulations become visible 
in the blue spectral regions, with emission components peaking at about 4020~\AA, 4600~\AA~and 5330~\AA~(rest wavelengths). 
These features are likely due to blends of Fe II features. 
Subsequent spectra (phases 13.2 d to 18.1 d) show other broad emission features peaking at about 6200 \AA~and 6450 \AA,
very likely line blends that become stronger with time. 

At later epochs (from 20.2 d to 30.4 d), higher S/N spectra were obtained using TNG and GTC. 
These spectra clearly show prominent He I lines (e.g. He I $\lambda$ 4471,  $\lambda$ 5876 and  $\lambda$ 7065) 
with broad P-Cygni profiles at an inferred velocity of 6000-7000 km s$^{-1}$ (as determined from the wavelength of 
the P-Cygni minimum).
A careful inspection of lower S/N spectra at earlier phases suggests that there was some evidence of broad 
He I lines at  epochs as early as +13.2 d.

From 20.2 d after maximum onwards, a very broad  emission feature peaking at 8400-8700 \AA~emerges, and becomes the strongest feature 
in the spectra of ASASSN-15ed. This feature is identified as the NIR Ca II triplet. Deblending the components of the
triplet, we obtain an ejection velocity of about 8500 km s$^{-1}$ for the gas where the lines form. 

We note that the NIR Ca II triplet is likely generated in the outermost ejecta, as this velocity is significantly 
larger than the bulk velocity deduced from the position of the broad He I absorption components.
The NIR Ca II blend is the most prominent spectral feature in our latest low S/N GTC spectrum, 
obtained 53.2 d after maximum light.

\subsection[]{Broad lines: comparison with SN~2010al} \label{spc_broad}

One of the most intriguing spectroscopic properties of ASASSN-15ed is the transformation from
almost featureless early-time spectra where only narrow P-Cygni profile lines of He I were visible, to the later spectra
dominated by very broad P-Cygni lines similar to those observed in SNe Ib (cf. Section \ref{spc_cfr}).
To our knowledge, a similar spectral metamorphosis has previously observed only in SN~2010al 
\citep{pasto15a} and in SN~2015G (Fraser et al. in preparation).

In Figure \ref{seqlines}, we show the temporal evolution of selected spectral features in a few representative
spectra of ASASSN-15ed (top panel) and SN~2010al (bottom panel). Approximately three weeks after maximum,
the spectra of ASASSN-15ed show a rather sharp transition from being dominated by narrow lines (v $\leq$ 1500 km~s$^{-1}$) to showing 
purely broad lines (v $\approx$ 6500 km~s$^{-1}$).
Although the relatively broad P-Cygni component is  weak in the early spectra of ASASSN-15ed, it is clearly
detected at those epochs and shows very little velocity evolution. The co-existence of two line components with
modest velocity evolution suggests that these features arise from different emitting regions:
 the broader He~I P-Cygni features are likely a signature of the
SN ejecta, while the narrow He~I P-Cygni lines are generated in the unperturbed, He-rich  CSM.
At early phases, the photosphere's location is within the dense shell, above which the narrow lines
form. The shell is either photoionized by early ejecta-CSM interaction in the inner CSM regions and/or by the initial shock breakout.
Once recombined, the shell becomes transparent and the underlying SN ejecta gradually emerge: in this phase, the spectrum
is dominated by broad P-Cygni lines.

\begin{figure}
\includegraphics[width=8.3cm,angle=0]{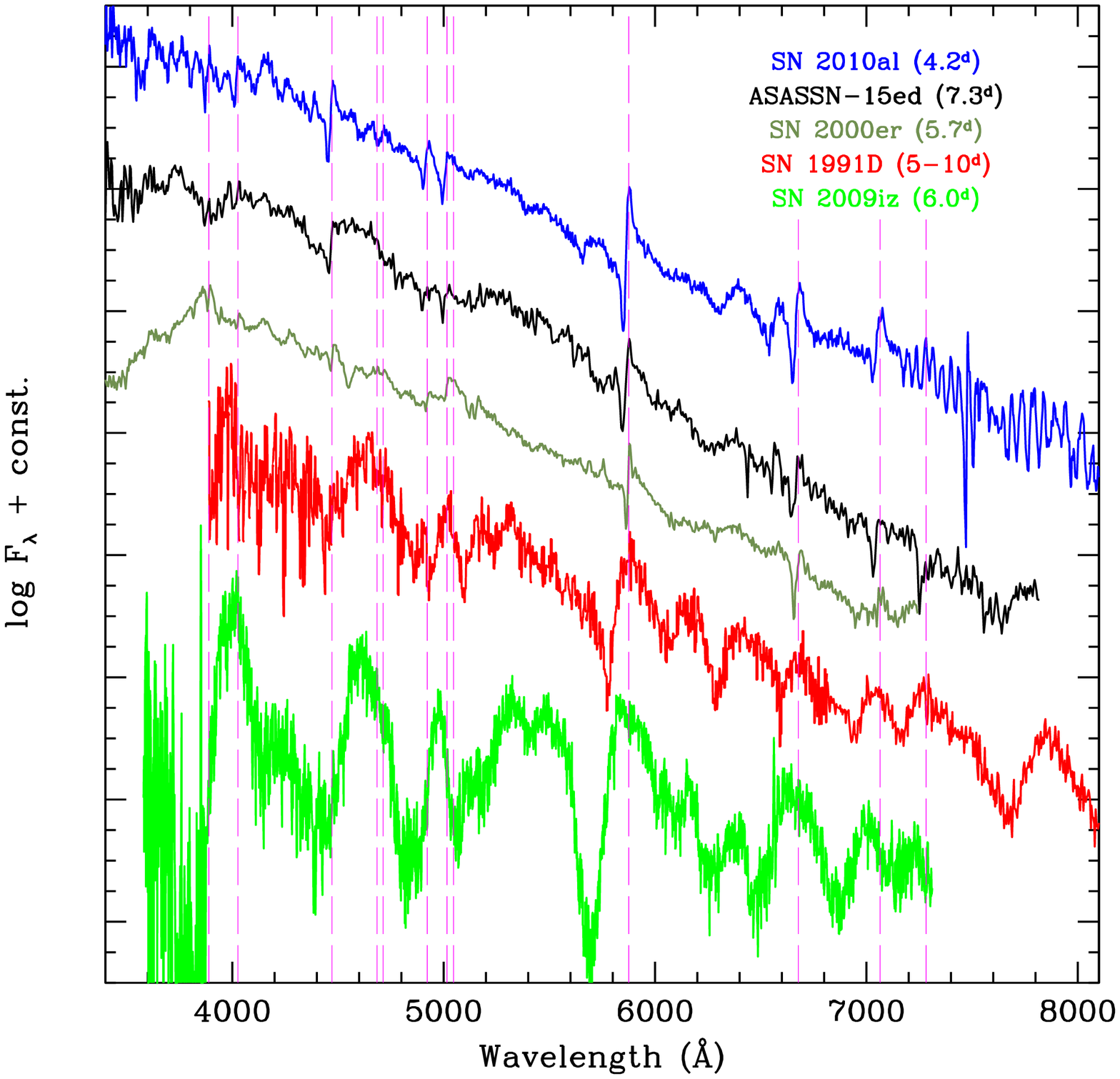} 
% \vspace{302pt}
 \caption{Comparison of an early-time spectrum of ASASSN-15ed obtained 7.3 d after maximum with spectra of the Type Ibn SNe
2010al and 2000er, the peculiar SN Ib 1991D and the typical SN Ib 2009iz at similar phases. All spectra have been Doppler and reddening corrected,
using the information available in the literature \protect\citep{ben02,pasto08a,pasto15a,mod14}. For SN 2009iz, the information on 
the reddening, redshift and epoch of the $V$-band maximum is taken from \citet{bia14}.
The phases of the five objects are relative to the maximum light. The dashed vertical lines mark the expected
positions of the He~I lines.\label{spec_cfr1}}
\end{figure}
 
SN~2010al shows a smooth transition, with spectral lines that become progressively broader with time and only one
velocity component is apparent observed at each epoch.
In the early spectra of SN~2010al, the narrow He~I P-Cygni lines are likely produced in a circumstellar shell, and indicate the presence
of gas moving at about 1000-1100 km s$^{-1}$.
With time, the He~I lines become progressively broader, reaching  v~$\sim$~1900-2300~km s$^{-1}$ at $\sim$20 d after maximum, and
5000-6000 km s$^{-1}$ at 1.5 months after maximum. This gradual transition may suggest the existence of slow-moving CSM which is 
shocked by the SN ejecta. However, this interpretation does not explain the presence of the broad blue-shifted absorption component.
An alternative possibility is the same scenario invoked for ASASSN-15ed, with the narrow line emitting region dominating the spectrum 
at early phases and the  broad line contribution from the SN ejecta emerging at later phase when the CSM becomes transparent.
In this case, the peculiar line velocity evolution would result from the convolved line profile of the two emitting 
regions, with their relative intensities that change with time.

\subsection[]{Comparison with Type Ib SN spectra} \label{spc_cfr}

In Figure \ref{spec_cfr1}, we compare an early-time spectrum of ASASSN-15ed (phase = 7.3 d) with spectra of two SNe Ibn
for which early-time observations are available  \citep[SNe~2000er and 2010al;][]{pasto08a,pasto15a}, with the peculiar Type Ib 
SN 1991D \citep{ben02}, and the normal SN Ib 2009iz \citep{mod14}. The reddening and redshift estimates adopted in the
comparison are those of the reference papers.
The spectra of SNe Ibn and SNe Ib are markedly different at early phases, since the He~I lines, with
P-Cygni profiles,  are very narrow (1000-1500 km s$^{-1}$) in Type Ibn SNe. As we have remarked in Section \ref{spc_ev}, the velocity
of these narrow components in SNe Ibn is indicative of the expansion velocity of the CSM. In addition, in early spectra of
SNe Ibn, features attributable to other ions are very weak.
In Type Ib SN spectra, we clearly see broad lines with P-Cygni profiles, indicating that the bulk of the material
ejected by the SN is moving with velocities of several thousands km s$^{-1}$, and we clearly identify lines of Fe II 
and intermediate-mass elements (IME), along with the prominent He~I features.

\begin{figure}
\includegraphics[width=8.3cm,angle=0]{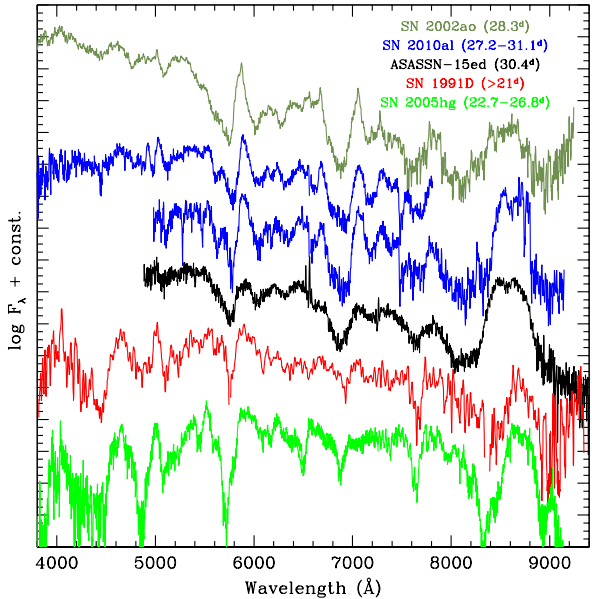} 
% \vspace{302pt}
 \caption{Comparison of a spectrum of ASASSN-15ed obtained 30.4 d after maximum with spectra of the Type Ibn SNe
2010al and 2002ao; the peculiar SN Ib 1991D and the typical SN Ib 2005hg at comparable phases. For SN~2005hg, the information on 
the reddening, redshift and epoch of the $V$-band maximum is taken from \citet{bia14}.
The spectrum of SN 2002ao \citep[taken from][]{pasto08a}
has been smoothed using a boxcar of 5 pixels, that of SN 1991D  \citep[from][]{mat01} using a boxcar of 7 pixels, and the +25.8 d spectrum of SN 2005hg
 \citep{mod14}  using a boxcar of 3 pixels.
The spectra have been Doppler and reddening corrected. The phases of the five objects (reported in brackets) are in days after the maximum light. \label{spec_cfr2}}
\end{figure}

In Figure \ref{spec_cfr2}, a relatively late GTC spectrum of ASASSN-15ed (phase = 30.4 d) is compared with roughly coeval spectra of
the Type Ibn SNe 2002ao and 2010al \citep{pasto08a,pasto15a}, the peculiar SN~Ib 1991D \citep{mat01} and the canonical SN~Ib 2005hg \citep{mod14}. 
At these later epochs there is more similarity between the spectra of our sample of SNe Ibn and the two SNe Ib.
Although in some SNe Ibn, the narrow He I lines are still visible at late phases \citep[e.g. in SN 2011hw,][]{smi12,pasto15a}, the
late spectra of SNe Ibn presented here have He I lines with intermediate to broad components that largely dominate over the narrow components.
The major difference among these spectra is that P-Cygni lines in SNe Ibn have  more symmetric profiles, 
while in SNe Ib the absorption dominates over the emission.
In addition, the shallow absorption components in the late spectra of SNe Ibn are broader than those of SNe Ib, suggesting 
a higher kinetic energy per mass unit in the SN ejecta for the former. 

In conclusion, although the spectral appearance of ASASSN-15ed (and, to a lesser extent, even  SN~2010al) changes with time, evolving from an
SN Ibn caught at early phases \citep[such as SN~2000er;][]{pasto08a} to a Type Ib SN,
some difference remains at late phases, with  the NIR Ca II triplet being remarkably stronger, and the P-Cygni profiles of the He~I lines
being broader in these two transitional SNe Ibn. Nonetheless, ASASSN-15ed and SN~2010al belong to a rare sub-group of SNe Ibn whose spectra show
a gradual transition to broad-line spectra of a Type Ib SN.

\section{Discussion and Summary} \label{ds}

In this paper, we have presented photometric and spectroscopic data of ASASSN-15ed, initially classified as a Type Ibn SN. However, the monitoring
campaign showed a major spectral evolution, with narrow He~I lines disappearing at late phases, and broad features
with P-Cygni profiles becoming more prominent with time (see Section \ref{spectroscopy}).
Although the spectroscopic evolution of ASASSN-15ed suggests a transitional case from a Type Ibn to a Type Ib SN, its light curve
is still markedly different from those of SNe Ib. In particular,
\begin{itemize}
\item With a peak magnitude of $M_r \approx -20$,  it is one of the most luminous SNe Ibn.
At maximum it is far more luminous than the average SNe Ib \citep[about $-17.5 \pm 0.3$ mag,][]{rich14},
although similarly bright SNe Ib do exist \citep{ben02,ner15}.
\item It has a very fast, almost linear photometric decline after maximum, faster than those of SNe Ib and
lasting about 50 d (see Table \ref{lcpar} and Figure \ref{fig_bolo}). 
\item During observed period, there is no evidence for the presence of a late-time light curve flattening on to
the $^{56}$Co decay tail. In fact, about two months post-maximum there is a remarkable increase in
the decline rate of the optical light curves, never previously observed -- to our knowledge -- in SNe Ib. 
This may either suggest that a very tiny amount of $^{56}$Ni was synthesized in the explosion or that
dust has formed in the SN environment, in analogy with SN~2006jc \citep{dica08,matt08,smi08}.
\end{itemize}

The evolution of ASASSN-15ed is somewhat reminiscent of the Type Ibn SN 2010al \citep{pasto15a}. The spectra of both objects 
were characterized by a blue continuum with superimposed narrow P-Cygni lines of He I during the initial phases. Later on, broad
P-Cygni components emerged for the He~I lines. In ASASSN-15ed, the velocities of these components remain roughly constant, while they increase with time in SN~2010al.
In both cases, this is interpreted as a signature of the SN ejecta. However, a photosphere receding through the ejecta is expected to produce
line velocities that decline with time. In  SN~2010al, the opposite trend of the velocity profile is puzzling. As discussed in Section \ref{spc_broad}, 
the presence of a circumstellar  shell may be an important ingredient to explain its evolution, with the velocity evolution representing the 
 combined contribution of two different line-forming regions. This would imply that, although SN~2010al and  ASASSN-15ed become spectroscopically 
similar to SNe Ib at later phases, the presence of a significant amount of H-depleted CSM has a key role in the evolution of these two objects. 
For this reason, we believe that their classification as transitioning Type Ibn/Ib SNe is well motivated, and supported
by the observations. 

What is still puzzling is the mechanism powering the early-time light
curve of  ASASSN-15ed. The initial shock breakout may contribute to the
light output at early phases. This process would also provide the
initial flash ionization of the thick CSM shell. However, given the
large radius\footnote{R$_{shell}$ $\approx$ 2 $\times$ 10$^4$ R$_\odot$, 
based on a temperature of 11000 K and a peak luminosity of 2 $\times$
10$^{43}$ erg s$^{-1}$.} of the thick shell, a quick recombination of this material has to
be delayed until the epochs at which the transition to a Type Ib is
observed. A continuous ionization of the shell could be maintained via
interaction of the SN ejecta with an inner circumstellar wind. With the
conversion of kinetic energy into thermal energy in the shock-heating
processes, this would also explain the high luminosity of ASSASN-15ed. 
Intermediate-width spectral lines typically arising from the shocked gas 
of the interacting SNe may also be reprocessed in the outer CSM, remaining unobserved 
as long as the shell is optically thick. Once the supply of ionizing radiation ceases
because the inner ejecta-CSM interaction decreases in strength (or stops
entirely), the shell recombines and the underlying Type Ib spectrum
becomes visible. Based on this scenario, we expect that once the SN
ejecta will reach the outer high-density circumstellar shell, a
flattening in the light curve of ASASSN-15ed will be observed.

Our attempt to link these transitioning SNe to a specific stellar progenitor is the ultimate goal in constraining their nature.
Although ASASSN-15ed and SN~2010al are very luminous at maximum, they do not have a luminous light curve at late times, and
we do not observe the classical $^{56}$Co decay tail.
This may be a consequence of significant dust formation\footnote{Unequivocal signatures of dust condensation were observed in SN~2006jc at phases above 50 d past maximum.
In SN~2010al, there was some
evidence of an NIR excess $-$ hence, possible dust condensation $-$  only at very late phases, above 200 d.}, though we cannot rule out that a 
very modest amount of  $^{56}$Ni is ejected in the explosion. In the latter case, the tiny $^{56}$Ni mass would link these transitioning SNe Ibn/Ib
to moderate-mass progenitors that had lost their mass via binary interaction, or to very massive stars that ejected
modest  $^{56}$Ni masses because of fall-back of stellar mantle material on to the core.
 
Unfortunately, the large distances of these SNe make the detection of the progenitor stars in pre-SN archive
images impossible. In addition, as the CSM has a predominant role in the SN evolution, an accurate modelling of the SN data 
and a consequent firm constraint on the parameters of the stellar precursor are problematic.

The pre-SN eruption observed before the explosion of SN~2006jc preferentially link SNe Ibn to massive WR stars \citep{fol07,pasto07,tom08}.
Subsequent studies \citep[see][and references therein]{pasto15d} proposed that most SNe Ibn come from WR progenitors, spanning the wide
range of observables from early WR with residual H mass \citep[Ofpe/WN9, e.g. SN~2011hw;][]{smi12} to He-poor WCO \citep[for SN~2006jc; see][]{tom08}, 
although alternative explosion channels -- though unpreferred -- cannot be definitely ruled out \citep{san13}.
The He-rich CSM seen in ASASSSN-15ed and SN 2010al could have been produced in pre-SN mass-loss events experienced by their WR progenitors.

We have observed so far a large diversity in SN Ibn, including double-peaked light curves \citep[iPTF13beo;][]{gor14},
late-time light curve flattening \citep[e.g. LSQ13ccw;][]{pasto15c}, and extremely fast post-peak luminosity declines \citep[e.g. LSQ13ccw;][]{pasto15b}.
ASASSN-15ed, which experienced a spectroscopic metamorphosis from a Type Ibn to a Type Ib SN, is a new and unprecedented contribution to the Type Ibn SN zoo.
Current and future SN surveys, such as ASAS-SN, will further increase the number of SN Ibn discoveries, including very nearby objects, which will hopefully
put more stringent constraints on the nature of their progenitors.

\section*{Acknowledgements}

We are grateful to S. Sim for insightful discussions.

AP, SB, NER, AH, LT, GT, and MT are partially supported by the PRIN-INAF 2014 with the project ``Transient Universe: unveiling new types of stellar explosions with PESSTO''. 
Support for JLP is in part by FONDECYT through the grant 1151445 and by the Ministry of Economy, Development, and Tourisms Millennium Science Initiative 
through grant IC120009, awarded to The Millennium Institute of Astrophysics, MAS.
NER acknowledges the support from the European Union Seventh Framework Programme (FP7/2007-2013) under grant agreement no. 267251 ``Astronomy Fellowships in Italy'' (AstroFIt). 
AMG acknowledges financial support by the Spanish Ministerio de Econom\'ia y Competitividad (MINECO) grant ESP2013-41268-R.
ST and UMN acknowledge support by TRR33 ``The Dark Universe'' of the German Research Foundation (DFG).
JFB is supported by NSF Grant PHY-1404311.
SD is supported by the ``Strategic Priority Research Program - The Emergence of Cosmological Structures'' of the Chinese Academy of
Sciences (grant no. XDB09000000).
TW-SH is supported by the DOE Computational Science Graduate Fellowship, grant number DE-FG02-97ER25308.
 EEOI is partially supported by the Brazilian agency CAPES (grant number 9229-13-2).
BJS is supported by NASA through Hubble Fellowship grant HF-51348.001 awarded by the Space Telescope Science Institute, which is operated by the 
Association of Universities for Research in Astronomy, Inc., for NASA, under contract NAS 5-26555.
SS and AW acknowledge support from NSF and NASA grants to ASU.
This work was supported by the Laboratory Directed Research and Development program at LANL.
We thank LCOGT and its staff for their continued support of ASAS-SN.

Development of ASAS-SN has been supported by NSF grant AST-0908816 and the Center for
Cosmology and AstroParticle Physics at the Ohio State University. 

This paper is based on observations made with the Italian Telescopio Nazionale Galileo
(TNG) operated on the island of La Palma by the Fundaci\'on Galileo Galilei of
the INAF (Istituto Nazionale di Astrofisica) and the 1.82-m Copernico Telescope of INAF-Asiago Observatory. This work also makes use of observations from the LCOGT network;  
it is also based on observations made with the Gran Telescopio Canarias (GTC), installed in the Spanish Observatorio del Roque de los Muchachos 
of the Instituto de Astrof\'isica de Canarias, in the Island of La Palma;
on observations made with the Nordic Optical Telescope (NOT), operated on the island of La Palma jointly by Denmark, Finland, Iceland,
Norway, and Sweden, in the Spanish Observatorio del Roque de los Muchachos of the Instituto de Astrof\'isica de Canarias; 
on observations obtained at the MDM Observatory, operated by Dartmouth College, Columbia University, Ohio State University, Ohio University, and the University of Michigan.
The LT is operated on the island of La Palma by Liverpool John Moores University in the Spanish Observatorio del 
Roque de los Muchachos of the Instituto de Astrofisica de Canarias with financial support from the UK Science and Technology Facilities Council.
Data presented in this paper are  also based on observations with the Fred Lawrence Whipple Observatory 1.5-m Telescope of SAO.
This paper uses data products produced by the OIR Telescope Data Center, supported by the Smithsonian Astrophysical Observatory.

This research has made use of the NASA/IPAC Extragalactic Database (NED) which is operated by the Jet Propulsion Laboratory, California Institute of Technology, under contract with the NASA. 
We acknowledge the usage of the HyperLeda data base (http://leda.univ-lyon1.fr).

\appendix

\section{Pseudo-bolometric light curve of ASASSN-15ed.}

In Table \ref{lcbol}, we present pseudo-bolometric data for ASASSN-15ed. We report in column 1 the phases from the $r$-band maximum,
while in column 2 the values of log $L$ for individual epochs, with associated uncertainties.

\begin{table}
\begin{center}
%\scriptsize
\caption{Pseudo-bolometric data for ASASSN-15ed. 
\label{lcbol} }
\begin{tabular}{cc}
\hline \hline
Phase & log $L$ \\ 
(days from peak) & (erg s$^{-1}$) \\ \hline
$-$12.96 & $<$42.85 \\
 $-$7.97 & $<$42.91 \\
 $-$3.64 & 43.232  0.066  \\
  0.56 & 43.336  0.067  \\
  3.61 & 43.236  0.066 \\ 
  5.55 & 43.155  0.066  \\
  6.23 & 43.112  0.065 \\ 
  7.25 & 43.042  0.064 \\ 
  7.29 & 43.032  0.064 \\ 
  8.23 & 42.992  0.065 \\ 
  8.55 & 42.976  0.065 \\ 
  9.22 & 42.957  0.064 \\ 
  9.25 & 42.953  0.064 \\ 
  9.57 & 42.932  0.064 \\ 
 10.60 & 42.877  0.064  \\
 11.25 & 42.836  0.063 \\ 
 13.23 & 42.730  0.065 \\ 
 16.51 & 42.564  0.065  \\
 18.11 & 42.475  0.069  \\
 18.46 & 42.456  0.067 \\ 
 19.46 & 42.400  0.067 \\ 
 22.28 & 42.321  0.063 \\ 
 22.39 & 42.317  0.063 \\ 
 22.39 & 42.313  0.064 \\ 
 24.18 & 42.234  0.074 \\ 
 25.20 & 42.193  0.069 \\ 
 26.18 & 42.154  0.068 \\ 
 30.21 & 42.038  0.068 \\ 
 30.35 & 42.032  0.069 \\ 
 33.33 & 41.942  0.070 \\ 
 35.26 & 41.880  0.069 \\ 
 37.21 & 41.812  0.070  \\
 37.27 & 41.803  0.069 \\ 
 40.22 & 41.676  0.070  \\
 44.14 & 41.573  0.069  \\
 48.25 & 41.459  0.069 \\ 
 50.21 & 41.394  0.068 \\ 
 52.14 & 41.285  0.070 \\ 
 53.26 & 41.174  0.069 \\ 
 55.11 & 40.990  0.076 \\ 
 64.26 & $<$40.73 \\
 66.28 & $<$40.66 \\
 \hline
\end{tabular}
\end{center}
\end{table}

\label{lastpage}

\end{document}